\documentclass{article}

\usepackage[dvipdfmx]{graphicx}
\usepackage{amsmath}        
\usepackage{amssymb}        
\usepackage{cite}
\usepackage[dvips]{color}
\usepackage{fancyhdr}

\def\erase#1{{}}
\def\EqArrerase#1{{}}

\makeatletter
 
 \@addtoreset{equation}{section}
\makeatother

\def\GL{{G\kern-.12em L\kern-.04em}}
\def\OSp{{O\kern-.11em S\kern-.04em p}}
\def\IOSp{{I\kern-.06em O\kern-.11em S\kern-.04em p}}
\def\MN{{M\kern-.14em N}}
\def\NM{{N\kern-.14em M}}
\def\NL{{N\kern-.14em L}}
\def\LN{{L\kern-.11em N}}
\def\ML{{M\kern-.14em L}}
\def\LM{{L\kern-.11em M}}
\def\RN{{R\kern-.11em N}}
\def\NR{{N\kern-.14em R}}
\def\RM{{R\kern-.11em M}}
\def\MR{{M\kern-.14em R}}
\def\RL{{R\kern-.11em L}}
\def\LR{{L\kern-.11em R}}
\def\RS{{R\kern-.11em S}}
\def\SR{{S\kern-.11em R}}
\def\SN{{S\kern-.11em N}}
\def\NS{{N\kern-.11em S}}
\def\SM{{S\kern-.11em M}}
\def\MS{{M\kern-.11em S}}
\def\SL{{S\kern-.11em L}}
\def\LS{{L\kern-.11em S}}
\def\sqr#1#2{{\vcenter{\hrule height.#2pt
      \hbox{\vrule width.#2pt height#1pt \kern#1pt
          \vrule width.#2pt}
      \hrule height.#2pt}}}
\def\bra0{\langle0|}
\def\ket0{|0\rangle}
\def\soeji#1_#2#3{#1_{#2}\cdots#1_{#3}}
\def\longgLRarrow{\longleftarrow\kern-3pt\relbar\kern-3pt\relbar\kern-3pt%
\longrightarrow}
\def\longLRarrow{\longleftarrow\kern-3pt\relbar\kern-3pt\longrightarrow}
\def\longLarrow{\longleftarrow\kern-3pt\relbar\kern-3pt\relbar\kern-3pt\relbar}
\def\longRarrow{\relbar\kern-3pt\relbar\kern-3pt\relbar\kern-3pt\longrightarrow}
\def\bothDer#1#2#3{%
\overset{\kern-.7em\stackrel{#1}{#2}}{\partial_{#3}}}
 
\makeatletter
 
 \@addtoreset{equation}{section}
\makeatother

\begin{document}
\thispagestyle{fancy}

\title{Vanishing Noether Current in Weyl Invariant Gravities}

\author{Ichiro Oda
\footnote{Electronic address: ioda@sci.u-ryukyu.ac.jp}
\\
{\it\small
\begin{tabular}{c}
Department of Physics, Faculty of Science, University of the 
           Ryukyus,\\
           Nishihara, Okinawa 903-0213, Japan\\      
\end{tabular}
}
}
\date{}

\maketitle

\thispagestyle{fancy}

\begin{abstract}

We revisit the issue that the Noether current associated with a local scale symmtery, or equivalently
the Weyl symmetry, identically vanishes. Based on only the second Noether theorem for a local symmetry, 
we prove that the Noether current associated with the Weyl symmetry is in general vanishing in any Weyl 
invariant gravitational theories in four dimensional Riemannian geometry. We also clarify the reason: The Weyl 
transformation is non-dynamical in the sense that it does not contain the derivative term of the transformation 
parameter as opposed to the conventional gauge transformation.  Finally, we apply this result to a quantum 
theory of a general Weyl invariant gravity and derive currents associated with choral symmetry, which is 
the $IOSp(10|10)$ supersymmetry.
 
\end{abstract}

\section{Introduction}

In recent years, we have watched a revival of interests in both globally and locally scale invariant
theories since a natural solution to the gauge hierarchy problem on the basis of supersymmetry
is at present under intense pressure from the LHC's exploration of physics at the TeV scale. 
If we drop the Higgs mass term from an action, the Standard Model is free from any mass
scale and it is therefore invariant under the scale transformations \cite{Bardeen}. It is
believed that when the primordial hot universe cools down to the temperature around 
the TeV scale, the electroweak phase transition occurs and the Higgs particle appears in the mass 
spectrum. At the energies above this phase transition the Standard Model would be invariant 
under the scale transformations. Moreoever, even if elementary particles have rest masses, 
they would become effectively massless at high energies owing to the huge kinetic energy
where the scale symmetries would become an exact symmetry.   
  
In almost all proposals based on the scale symmetries, they pursue the possibility that the Standard
Model could survive modulo some minor modifications all the way to the Planck scale where we cannot 
ignore contributions from quantum gravity. When we take quantum effects associated with gravity
into consideration, it is not a global but a local scale symmetry that we have to respect 
since the presence of black holes enables many of global scale symmetries to break. For instance, 
when a black hole evaporates at the quantum level, the baryon and lepton numbers are not conserved 
whereas gauge quantum numbers such as electric and magnetic charges are precisely conserved 
since they are measured by the flux integrals at infinity. Thus, if we want to advocate the proposals 
based on the scale symmetries, we have to construct a standard model within the framework of 
a Weyl invariant or a locally scale invariant gravity. 

Several years ago, in Ref. \cite{Jackiw} it was pointed out that the Noether current associated with 
the Weyl symmetry in a Weyl invariant scalar-tensor gravity \cite{Fujii} vanishes identically by both
a direct evaluation and the second Noether theorem \cite{Noether, Bak}, and afterwards it was shown that 
the same result holds for conformal gravity by dealing with it as a gauge natural theory \cite{Campigotto}, 
the unimodular gravity \cite{Oda-U} and the Weyl transverse gravity \cite{Oda-C, Alonso} by a direct
calculation.  

The purposes of this article are threefold: First, we prove that the Noether current associated with
the Weyl symmetry is vanishing in any Weyl invariant gravities in four dimensional Riemannian geometry.
Second, we make use of only the second Noether theorem \cite{Noether, Bak} in our proof. In case of 
conformal gravity, the proof was carried out by formulating conformal gravity as a gauge natural theory \cite{Campigotto}
which is not familiar to many of researchers, so it is useful to present a new and general proof based on the
second Noether theorem. As a bonus of this new proof, we can clarify the reason why the Noether current 
of the Weyl symmetry is vanishing. It is usually considered that the vanishing Noether current associated with 
the Weyl symmetry reflects the fact that the Weyl symmetry does not have any dynamical role \cite{Jackiw}. 
However, in a Weyl invariant scalar-tensor gravity a ghost-like scalar field, which is sometimes called ``spurion'', 
is indeed a dynamical field with the kinetic term
and appears in the mass spectrum, in particular, except for the unitary gauge, even if it is confined in the unphysical 
sector via the BRST quartet mechanism \cite{Oda-Q}. We show that the true reason of the vanishing Noether
current is that the Weyl transformation is non-dynamical in the sense that it does not contain 
the derivative term of the transformation parameter as opposed to the conventional gauge transformation.   
Finally, we apply the vanishing Weyl Noether current to a quantum theory of a general Weyl invariant gravity
and show that there is a huge global symmetry called ``choral symmetry'' in the quantum gravity \cite{Nakanishi, N-O-text}
owing to the absence of the vanishing Noether current of the Weyl symmetry. 

The paper is organized as follows. In Section 2, we present a second Noether theorem for Lagrangians
that depend at most on second derivatives of dynamical fields since the Riemannian tensors contain
at most second derivatives of the metric tensor. In Section 3, in order to prove the vanishing Weyl 
Noether current the second theorem is utilized for any Weyl invariant gravities which include Weyl invariant 
scalar-tensor gravities and conformal gravity. In Section 4, we apply the vanishing Noether current of 
the Weyl symmetry to a quantum theory of a general Weyl invariant gravity and obtain the currents for 
choral symmetry, i.e., the $IOSp(10|10)$ supersymmetry. The final section is devoted to discussion.

\section{Noether current of local symmetry}

In this section, we wish to show that the Noether current associated with a local
symmetry can be always written as a form which is identically conserved. This
theorem has been proved by Noether \cite{Noether} and the explicit form of the
current has been given in Ref. \cite{Bak}. For later convenience, we slightly generalize
the formalism in Ref. \cite{Bak} to Lagrangians that depend at most on second derivatives of 
dynamical fields since the Riemannian tensors contain at most second derivatives of the metric tensor.

Let us start with an action $S$ which depends at most on second derivatives of a generic dynamical 
field $\phi_i$ where the index $i$ labels the species of fields, and a local gauge transformation 
$\delta \phi_i$ containing at most first derivative of the gauge transformation parameter 
$\theta (x)$:\footnote{We follow the notation and conventions of MTW textbook \cite{MTW}. 
The Minkowski metric tensor is denoted by $\eta_{\mu\nu}$; $\eta_{00} = - \eta_{11} = - \eta_{22} 
= - \eta_{33} = -1$ and $\eta_{\mu\nu} = 0$ for $\mu \neq \nu$. The Riemann curvature tensor 
and the Ricci tensor are respectively defined by $R^\rho{}_{\sigma\mu\nu} = \partial_\mu \Gamma^\rho_{\sigma\nu} 
- \partial_\nu \Gamma^\rho_{\sigma\mu} + \Gamma^\rho_{\lambda\mu} \Gamma^\lambda_{\sigma\nu} 
- \Gamma^\rho_{\lambda\nu} \Gamma^\lambda_{\sigma\mu}$ and $R_{\mu\nu} = R^\rho{}_{\mu\rho\nu}$.} 
\begin{eqnarray}
S = \int d^4 x \, {\cal L} ( \phi_i, \partial \phi_i, \partial^2 \phi_i ),
\label{Action}  
\end{eqnarray}
which is invariant under an infinitesimal local gauge transformation:
\begin{eqnarray}
\delta \phi_i = \Delta_i (\phi) \theta (x) + \Delta_i^\mu (\phi) \partial_\mu \theta (x),
\label{Gauge transf}  
\end{eqnarray}
where $\Delta_i$ and $\Delta_i^\mu$ may depend on $\phi_i$. 

Let us first derive the Noether identity and the field equation. Taking a variation of the action 
(\ref{Action}) under the local transformation (\ref{Gauge transf}) leads to
\begin{eqnarray}
\delta S &=& \int d^4 x \, \left( \frac{\partial {\cal L}}{\partial \phi_i} \delta \phi_i 
+ \frac{\partial {\cal L}}{\partial \partial_\mu \phi_i} \partial_\mu \delta \phi_i
+ \frac{\partial {\cal L}}{\partial \partial_\mu \partial_\nu \phi_i} \partial_\mu \partial_\nu 
\delta \phi_i \right)
\nonumber\\ 
&=& \int d^4 x \, \left[ \frac{\delta {\cal L}}{\delta \phi_i} \Delta_i 
- \partial_\mu \left( \frac{\delta {\cal L}}{\delta \phi_i} \Delta_i^\mu \right) \right] \theta,
\label{Var-Action}  
\end{eqnarray}
where we have performed the integration by parts and $\frac{\delta {\cal L}}{\delta \phi_i}$
denotes the following Euler-Lagrange derivative:
\begin{eqnarray}
\frac{\delta {\cal L}}{\delta \phi_i} = \frac{\partial {\cal L}}{\partial \phi_i}
- \partial_\mu \frac{\partial {\cal L}}{\partial \partial_\mu \phi_i}
+ \partial_\mu \partial_\nu \frac{\partial {\cal L}}{\partial \partial_\mu \partial_\nu \phi_i}.
\label{EL-der}  
\end{eqnarray}
As a result, we have the Noether identity:
\begin{eqnarray}
\frac{\delta {\cal L}}{\delta \phi_i} \Delta_i 
- \partial_\mu \left( \frac{\delta {\cal L}}{\delta \phi_i} \Delta_i^\mu \right) = 0.
\label{Noe-id}  
\end{eqnarray}
Similarly, considering an arbitrary variation of the action, the field equation reads
\begin{eqnarray}
\frac{\delta {\cal L}}{\delta \phi_i} = 0.
\label{Field-eq}  
\end{eqnarray}
 
Next, we find that without recourse to the field equation, the variation of the Lagrangian
under the local transformation (\ref{Gauge transf}) takes the form:
\begin{eqnarray}
\delta {\cal L} &=& \frac{\partial {\cal L}}{\partial \phi_i} \delta \phi_i 
+ \frac{\partial {\cal L}}{\partial \partial_\mu \phi_i} \partial_\mu \delta \phi_i
+ \frac{\partial {\cal L}}{\partial \partial_\mu \partial_\nu \phi_i} \partial_\mu \partial_\nu \delta \phi_i
\nonumber\\
&=& M \theta + M^\mu \partial_\mu \theta + M^{\mu\nu} \partial_\mu \partial_\nu \theta
+ M^{\mu\nu\rho} \partial_\mu \partial_\nu \partial_\rho \theta,
\label{Var-Lag}  
\end{eqnarray}
where we have defined
\begin{eqnarray}
M &=& \frac{\partial {\cal L}}{\partial \phi_i} \Delta_i 
+ \frac{\partial {\cal L}}{\partial \partial_\mu \phi_i} \partial_\mu \Delta_i
+ \frac{\partial {\cal L}}{\partial \partial_\mu \partial_\nu \phi_i} \partial_\mu \partial_\nu \Delta_i,
\nonumber\\
M^\mu &=& \frac{\partial {\cal L}}{\partial \phi_i} \Delta_i^\mu 
+ \frac{\partial {\cal L}}{\partial \partial_\nu \phi_i} ( \Delta_i \delta_\nu^\mu + \partial_\nu \Delta_i^\mu )
+ \frac{\partial {\cal L}}{\partial \partial_\nu \partial_\rho \phi_i} ( 2 \partial_\nu \Delta_i \delta_\rho^\mu 
+ \partial_\nu \partial_\rho \Delta_i^\mu ),
\nonumber\\
M^{\mu\nu} &=& \frac{\partial {\cal L}}{\partial \partial_\mu \phi_i} \Delta_i^\nu 
+ \frac{\partial {\cal L}}{\partial \partial_\mu \partial_\rho \phi_i} ( \Delta_i \delta_\rho^\nu 
+ 2 \partial_\rho \Delta_i^\nu ),
\nonumber\\
M^{\mu\nu\rho} &=& M^{\nu\mu\rho} = \frac{\partial {\cal L}}{\partial \partial_\mu \partial_\nu \phi_i} 
\Delta_i^\rho. 
\label{M}  
\end{eqnarray}
With the help of the Noether identity (\ref{Noe-id}), it turns out that $M$ can be expressed 
in terms of $M^\mu, M^{\mu\nu}$ and $M^{\mu\nu\rho}$ as
\begin{eqnarray}
M = \partial_\mu M^\mu - \partial_\mu \partial_\nu M^{(\mu\nu)} 
+ \partial_\mu \partial_\nu \partial_\rho M^{(\mu\nu\rho)}, 
\label{M-rel}  
\end{eqnarray}
where we have introduced the symmetrization notation such as $M^{(\mu\nu)}
\equiv \frac{1}{2} ( M^{\mu\nu} + M^{\nu\mu} )$. (Later, we will also adopt the 
antisymmetrization notation such as $M^{[\mu\nu]} \equiv \frac{1}{2} ( M^{\mu\nu} - M^{\nu\mu} )$.)
Then, since the action (\ref{Action}) is invariant under the local gauge transformation 
(\ref{Gauge transf}), by means of Eq. (\ref{M-rel}) the variation of the Lagrangian $\delta {\cal L}$ 
in (\ref{Var-Lag}) can be cast to the form of a total derivative:
\begin{eqnarray}
\delta {\cal L} &=& \partial_\mu \Bigl[ M^\mu \theta + M^{(\mu\nu)} \partial_\nu \theta
- \partial_\nu M^{(\mu\nu)} \theta + \partial_\nu \partial_\rho M^{(\mu\nu\rho)} \theta 
\nonumber\\
&-& \partial_\rho M^{(\mu\nu\rho)} \partial_\nu \theta
+ M^{(\mu\nu\rho)} \partial_\nu \partial_\rho \theta \Bigr]
\nonumber\\
&\equiv& \partial_\mu L^\mu.
\label{Var-Lag-X}  
\end{eqnarray}

Now let us rewrite the variation $\delta {\cal L}$ in (\ref{Var-Lag}) by using the field equation
(\ref{Field-eq}), thereby reaching an alternate divergence formula:
\begin{eqnarray}
\delta {\cal L} &=& \partial_\mu \Bigl( \frac{\partial {\cal L}}{\partial \partial_\mu \phi_i} \delta \phi_i 
- \partial_\nu \frac{\partial {\cal L}}{\partial \partial_\mu \partial_\nu \phi_i} \delta \phi_i
+ \frac{\partial {\cal L}}{\partial \partial_\mu \partial_\nu \phi_i} \partial_\nu \delta \phi_i \Bigr)
\nonumber\\
&\equiv& \partial_\mu K^\mu.
\label{Var-Lag-K}  
\end{eqnarray}
From the two equations (\ref{Var-Lag-X}) and (\ref{Var-Lag-K}), we can define the Noether current
\begin{eqnarray}
J^\mu = K^\mu - L^\mu,
\label{Current-J}  
\end{eqnarray}
which is obviously conserved:
\begin{eqnarray}
\partial_\mu J^\mu = 0.
\label{Current-cons}  
\end{eqnarray}
A straightforward calculation of the Noether current $J^\mu$ yields the desired result:
\begin{eqnarray}
J^\mu = \partial_\nu \Bigl[ M^{[\mu\nu]} \theta + \frac{2}{3} \partial_\rho \left( M^{\mu[\nu\rho]} 
- M^{\nu[\mu\rho]} - M^{\rho[\mu\nu]} \right) \theta 
+ \frac{2}{3} M^{\rho[\mu\nu]} \partial_\rho \theta \Bigr],
\label{J-exp}  
\end{eqnarray}
which is in general nonvanishing and is identically conserved.

As an important remark, as can be understood in the above derivation of the Noether 
current, when an additional total derivative term is added to the original 
action in (\ref{Action}), the expression of the Noether current (\ref{J-exp}) remains unchanged 
although each expression of $L^\mu$ and $K^\mu$ might be modified. This is because a total 
derivative term is always invariant under a local transformation.

\section{Vanishing Noether current}

In this section, based on the formula of the Noether current (\ref{J-exp}) for the local
transformation, we will prove that the Noether current associated with the Weyl 
transformation is identically vanishing in any Weyl invariant gravities in four space-time dimensions.

First, let us discuss conformal gravity, which has the well-known Lagrangian:
\begin{eqnarray}
{\cal L}_{cg} (g) = \alpha \sqrt{-g} \, C_{\mu\nu\rho\sigma} C^{\mu\nu\rho\sigma},
\label{CG}  
\end{eqnarray}
where $\alpha$ is a dimensionless coupling constant and $C_{\mu\nu\rho\sigma}$ is 
the conformal tensor defined as
\begin{eqnarray}
C_{\mu\nu\rho\sigma} &=& R_{\mu\nu\rho\sigma} - \frac{1}{2} ( g_{\mu\rho} R_{\nu\sigma}
- g_{\mu\sigma} R_{\nu\rho} - g_{\nu\rho} R_{\mu\sigma} + g_{\nu\sigma} R_{\mu\rho} )
\nonumber\\
&+& \frac{1}{6} ( g_{\mu\rho} g_{\nu\sigma} - g_{\mu\sigma} g_{\nu\rho} ) R.
\label{C-tensor}  
\end{eqnarray}
The Riemann curvature tensor $R_{\mu\nu\rho\sigma}$ (and its contractions) is a function
of $g, \partial g, \partial^2 g$, so is the conformal tensor. Thus, the Lagrangian (\ref{CG})
is also a function of $g, \partial g, \partial^2 g$, so we can apply the result (\ref{J-exp})
to the case of conformal gravity. 

The Weyl transformation, or a local scale transformation, is defined 
as\footnote{For later convenience, we have added the Weyl transformation for a real scalar
field $\phi$.}
\begin{eqnarray}
\delta g_{\mu\nu} = 2 \theta (x) g_{\mu\nu}, \qquad
\delta \phi = - \theta (x) \phi,
\label{Weyl-transf}  
\end{eqnarray}
where $\theta (x)$ is the infinitesimal transformation parameter. 
The key observation is that in the Weyl transformation there is no derivative term
of the infinitesimal transformation parameter $\theta (x)$, which should be contrasted
with the conventional gauge transformation where we have such a derivative term 
in the transformation law like $\delta A_\mu = \partial_\mu \theta$. Consequently, we have 
$\Delta_i^\mu = 0$ in Eq. (\ref{Gauge transf}), which gives us    
\begin{eqnarray}
M^{[\mu\nu]} = 0, \qquad
M^{\mu\nu\rho} = 0.
\label{Van-M}  
\end{eqnarray}
Accordingly, from Eq. (\ref{J-exp}), the Noether current assoiciated with the Weyl 
transformation in conformal gravity identically vanishes:
\begin{eqnarray}
J^\mu = 0.
\label{CG-V}  
\end{eqnarray}
Incidentally, this result was previously obtained by formulating conformal gravity as a gauge
natural theory and carrying out an intricated calculation \cite{Campigotto}. It is worthwhile to emphasize 
that our derivation not only is more general and concise than that in Ref. \cite{Campigotto} but also
clearly accounts for the reason why the Noether current of the Weyl symmetry is identically vanishing; 
the Weyl transformation does not include the derivative term in the transformation law.

Next, let us discuss a Weyl invariant scalar-tensor gravity whose Lagrangian is of form:
\begin{eqnarray}
{\cal L}_{st} (g, \phi) = \sqrt{-g} \left( \frac{1}{12} \phi^2 R + \frac{1}{2} g^{\mu\nu}
\partial_\mu \phi \partial_\nu \phi \right).
\label{Weyl-st}  
\end{eqnarray}
Recall that this Lagrangian is obtained from the Einstein-Hilbert Lagrangian of general relativity
via a replacement of the metric tensor as
\begin{eqnarray}
\tilde g_{\mu\nu} = \phi^2 g_{\mu\nu}.
\label{Tilde-g}  
\end{eqnarray}
Actually, starting with the Einstein-Hilbert Lagrangian we can have the Lagrangian (\ref{Weyl-st})
up to a total derivative term:
\begin{eqnarray}
{\cal L}_{EH} (\tilde g) &\equiv& \frac{1}{12} \sqrt{- \tilde g} R (\tilde g)
\nonumber\\
&=& \frac{1}{12} \sqrt{-g} ( \phi^2 R - 6 \phi \Box \phi )
\nonumber\\
&=& \sqrt{-g} \left( \frac{1}{12} \phi^2 R + \frac{1}{2} g^{\mu\nu}
\partial_\mu \phi \partial_\nu \phi \right) - \frac{1}{2} \partial_\mu ( \sqrt{-g} g^{\mu\nu} 
\phi \partial_\nu \phi )
\nonumber\\
&=& {\cal L}_{st} (g, \phi) - \partial_\mu I^\mu (g, \phi),
\label{EH-Lag}  
\end{eqnarray}
where 
\begin{eqnarray}
I^\mu (g, \phi) \equiv \frac{1}{2} \sqrt{-g} g^{\mu\nu} \phi \partial_\nu \phi.
\label{Def-I}  
\end{eqnarray}

In the case at hand, we can present two kinds of proofs that the Noether current 
associated with the Weyl symmetry is identically vanishing, one of which is that
since the composite metric tensor $\tilde g_{\mu\nu}$ in (\ref{Tilde-g}) is manifestly invariant
under the Weyl transformation (\ref{Weyl-transf}), we have no longer the Weyl
symmetry in the Lagrangian ${\cal L}_{EH} (\tilde g)$. The formula of the Noether
current does not depend on total derivative terms in a classical Lagrangian as
remarked in the previous section, so the Noether current of the Weyl symmetry in
the Lagrangian ${\cal L}_{st} (g, \phi)$ is identically vanishing. 

The other proof follows the same logic as in conformal gravity.  To do that, let us
begin with 
\begin{eqnarray}
{\cal L}_{st} (g, \phi)^\prime \equiv \frac{1}{12} \sqrt{-g} ( \phi^2 R - 6 \phi \Box \phi ),
\label{Mod-Lag}  
\end{eqnarray}
which is equivalent to the Lagrangian of the Weyl invariant scalar-tensor gravity ${\cal L}_{st} 
(g, \phi)$ up to a surface term as seen in Eq. (\ref{EH-Lag}) and is invariant under the Weyl
transformation (\ref{Weyl-transf}). Note that the Lagrangian (\ref{Mod-Lag}) is a function
of at most second derivatives of $g_{\mu\nu}$ and $\phi$. Thus, we can apply the formula of 
the Noether current (\ref{J-exp}) to this Lagrangian as well and obtain the vanishing Noether current
for the Weyl symmetry in the both Lagrangians ${\cal L}_{st} (g, \phi)$ and
${\cal L}_{st} (g, \phi)^\prime$. Of course, as can be verified by an explicit calculation, 
although the Noether current $J^\mu$ is identically vanishing in ${\cal L}_{st} (g, \phi)$ and
${\cal L}_{st} (g, \phi)^\prime$, it does not always mean that both $L^\mu$ and $K^\mu$
are identically vanishing. In fact, in the Lagrangian ${\cal L}_{st} (g, \phi)^\prime$ we have
$L^\mu = K^\mu =0$ whereas in the Lagrangian ${\cal L}_{st} (g, \phi)$ we have
$L^\mu = K^\mu = - \frac{1}{2} \sqrt{-g} g^{\mu\nu} \phi^2 \partial_\nu \theta$ \cite{Jackiw}.    

To close this section, it is worthwhile to mention that in four space-time dimensions there are 
only two classes of gravitational theories which are invariant under the Weyl transformation 
in the Riemannian geometry, those are, conformal gravity and Weyl invariant scalar-tensor
gravity. In particular, the latter theory can be constructed out of a pure metric Lagrangian 
via the replacement of the metric tensor $g_{\mu\nu}$ by $\tilde g_{\mu\nu}$ in 
Eq. (\ref{Tilde-g}).\footnote{For instance, starting with a Lagrangian of the $R^2$ gravity,
${\cal L}_{R^2} = \sqrt{-g} \, R^2$, we can make the corresponding Weyl invariant scalar-tensor
gravity with a Lagrangian $\tilde {\cal L}_{R^2} \equiv \sqrt{- \tilde g} R^2 (\tilde g) 
= \sqrt{-g} \, ( R - 6 \phi^{-1} \Box \phi )^2$.} 
In fact, it is fairly easy to show that this is the only method to construct a Weyl invariant
scalar-tensor gravity \cite{Tsamis}. As a result, it can be concluded that the Noether current
associated with the Weyl symmetry is identically vanishing in all the Weyl invariant gravities 
in the four-dimensional Riemannian geometry.

\section{Application of vanishing Weyl current to quantum gravity}
 
In this section, we present an application of the result obtained so far to a quantum theory of
a Weyl invariant gravity in four dimensions.    

Let us start with a general Weyl invariant gravity whose Lagrangian is a function of at most 
second derivatives of fields:
\begin{eqnarray}
{\cal L}_c = \sqrt{-g} \left( \frac{1}{12} \phi^2 R + \frac{1}{2} g^{\mu\nu} \partial_\mu \phi \partial_\nu \phi 
+ \alpha C_{\mu\nu\rho\sigma} C^{\mu\nu\rho\sigma} \right).
\label{WI-gravity}  
\end{eqnarray}
As a gauge condition for the general coordinate transformation (GCT), we adopt ``the extended 
de Donder gauge'', which is defined by \cite{Oda-Q}
\begin{eqnarray}
\partial_\mu ( \sqrt{-g} g^{\mu\nu} \phi^2 ) = 0.
\label{Ext-de-Donder}  
\end{eqnarray}
which is also invariant under the Weyl transformation (\ref{Weyl-transf}). 
Moreover, we take ``the scalar gauge condition'' for the Weyl transformation \cite{Oda-Q}: 
\begin{eqnarray}
\partial_\mu ( \sqrt{-g} g^{\mu\nu} \phi \partial_\nu \phi ) = 0,
\label{Scalar-gauge}  
\end{eqnarray}
which can be alternatively written as 
\begin{eqnarray}
\Box \, \phi^2 = 0.
\label{Alt-Scalar-gauge}  
\end{eqnarray}

After taking these gauge conditions, the gauge-fixed and BRST invariant quantum Lagrangian reads:
\begin{eqnarray}
{\cal L}_q &=& \sqrt{-g} \left( \frac{1}{12} \phi^2 R + \frac{1}{2} g^{\mu\nu} \partial_\mu \phi \partial_\nu \phi 
+ \alpha C_{\mu\nu\rho\sigma} C^{\mu\nu\rho\sigma} \right)
+ \sqrt{-g} g^{\mu\nu} \phi \partial_\mu B \partial_\nu \phi
\nonumber\\
&-& \sqrt{-g} g^{\mu\nu} \phi^2 ( \partial_\mu b_\nu + i \partial_\mu \bar c_\lambda  \partial_\nu c^\lambda )
 - i \sqrt{-g} g^{\mu\nu} \phi^2 \partial_\mu \bar c 
\partial_\nu c,
\label{WI-q-Lag}  
\end{eqnarray}
which can be rewritten as
\begin{eqnarray}
{\cal L}_q = \sqrt{-g} \left( \frac{1}{12} \phi^2 R + \alpha C_{\mu\nu\rho\sigma} C^{\mu\nu\rho\sigma} \right)
- \frac{1}{2} \sqrt{-g} g^{\mu\nu} E_{\mu\nu},
\label{WI-q-Lag2}  
\end{eqnarray}
where we have defined  
\begin{eqnarray}
E_{\mu\nu} &\equiv& - \frac{1}{2} \partial_\mu \phi \partial_\nu \phi + \phi^2 ( \partial_\mu b_\nu 
+ i \partial_\mu \bar c_\lambda  \partial_\nu c^\lambda )
\nonumber\\
&-& \phi \partial_\mu B \partial_\nu \phi + i \phi^2 \partial_\mu \bar c \partial_\nu c
+ ( \mu \leftrightarrow \nu ). 
\label{E}  
\end{eqnarray}
Here $b_\mu$ and $B$ are the Nakanishi-Lautrup fields, and $c^\mu, \bar c_\mu, c$ and $\bar c$ are 
the Faddeev-Popov (FP) ghosts.

We would like to rewrite the Lagrangian (\ref{WI-q-Lag2}) further into a more suggestive form.
For this purpose, we introduce ``the dilaton'' $\sigma (x)$ by defining
\begin{eqnarray}
\phi (x) \equiv e^{\sigma (x)}.
\label{Dilaton}  
\end{eqnarray}
Then, setting $X^M = \{ x^\mu, b_\mu, \sigma, B, c^\mu, \bar c_\mu, c, \bar c \}$, 
the Lagrangian (\ref{WI-q-Lag2}) is recast to the form:
\begin{eqnarray}
&{}& {\cal L}_q = \sqrt{-g} \left( \frac{1}{12} \phi^2 R + \alpha C_{\mu\nu\rho\sigma} C^{\mu\nu\rho\sigma} \right)
- \frac{1}{2} \sqrt{-g} g^{\mu\nu} \phi^2 \eta_{NM} \partial_\mu X^M \partial_\nu X^N
\nonumber\\
&=& \sqrt{-g} \left( \frac{1}{12} \phi^2 R + \alpha C_{\mu\nu\rho\sigma} C^{\mu\nu\rho\sigma} \right)
- \frac{1}{2} \sqrt{-g} g^{\mu\nu} \phi^2 \partial_\mu X^M \tilde \eta_{MN} \partial_\nu X^N,
\label{Choral-OSp-Lag}  
\end{eqnarray}
where we have introduced an $IOSp(10|10)$ metric $\eta_{NM} = \eta_{MN}^T \equiv \tilde \eta_{MN}$ 
defined as \cite{Oda-Q}
\begin{eqnarray}
\eta_{NM} = \tilde \eta_{MN} =
\left(
\begin{array}{cc|cc|cc|cc}
     &                \delta_\mu^\nu &     &   &     &  \\ 
\delta^\mu_\nu  &                    &    &    &    &   \\ 
\hline
    &        &               -1    &   -1      &     &    &      \\ 
    &        &               -1    &  0       &       &    &    \\
\hline   
    &        &     &    &       &   -i\delta_\mu^\nu  &   & \\  
    &        &     &    &   i\delta^\mu_\nu &   &    & \\
\hline
    &        &                &        &       &     &         &  -i \\  
    &        &                &        &       &     &          i     & \\
\end{array}
\right)_.
\label{OSp-metric}  
\end{eqnarray}

It turns out that from the field equations $X^M$ obeys the very simple equation:
\begin{eqnarray}
g^{\mu\nu} \partial_\mu \partial_\nu X^M = 0.
\label{X-M-eq}  
\end{eqnarray}
Together with the gauge condition $\partial_\mu ( \sqrt{-g} g^{\mu\nu} \phi^2 ) = 0$,
Eq. (\ref{X-M-eq}) produces the two kinds of conserved currents:
\begin{eqnarray}
{\cal P}^{\mu M} &\equiv& \sqrt{-g} g^{\mu\nu} \phi^2 \partial_\nu X^M 
= \sqrt{-g} g^{\mu\nu} \phi^2 \bigl( 1 \overset{\leftrightarrow}{\partial}_\nu X^M \bigr),
\nonumber\\
{\cal M}^{\mu M N} &\equiv& \sqrt{-g} g^{\mu\nu} \phi^2 \bigl( X^M 
\overset{\leftrightarrow}{\partial}_\nu Y^N \bigr),
\label{Cons-currents}  
\end{eqnarray}
where we have defined $X^M \overset{\leftrightarrow}{\partial}_\mu Y^N \equiv
X^M \partial_\mu Y^N - ( \partial_\mu X^M ) Y^N$. 

Actually, we can show that there exists an $IOSp(10|10)$ supersymmetry as a global symmetry
which is generated by the currents in (\ref{Cons-currents}). The infinitesimal $OSp$ rotation is defined by
\cite{Oda-Q}
\begin{eqnarray}
\delta X^M = \eta^{ML} \varepsilon_{LN} X^N \equiv \varepsilon^M{}_N X^N,
\label{OSp-rot}  
\end{eqnarray}
where $\eta^{MN}$ is the inverse matrix of $\eta_{MN}$.
In order to find the conserved current, we assume that the infinitesimal parameter $\varepsilon_{MN}$ 
depends on the space-time coordinates $x^\mu$, i.e., $\varepsilon_{MN} = \varepsilon_{MN} (x^\mu)$.
It is worth noticing that since $X^M$ contains the dilaton $\sigma$ among its components, 
the $OSp$ rotation (\ref{OSp-rot}) induces a local translation for the dilaton $\sigma$:
\begin{eqnarray}
\delta \sigma = \eta^{\sigma L} \varepsilon_{LN} X^N
= \eta^{\sigma B} \varepsilon_{BN} X^N
= - \varepsilon_{BN} X^N
= - \epsilon (x),
\label{Dilaton-Var}  
\end{eqnarray}
where we have defined $\epsilon (x) \equiv \varepsilon_{BN} X^N$ and used 
\begin{eqnarray} 
\begin{pmatrix}
   -1 & -1 \\
   -1 & 0
\end{pmatrix}^{-1}
= \begin{pmatrix}
   0 & -1 \\
   -1 & 1
\end{pmatrix}.
\label{Matrix}  
\end{eqnarray}
Thus, in order to compensate for some inessential terms, at the same time we have to perform 
a Weyl transformation:
\begin{eqnarray}
\delta \phi = - \epsilon (x) \phi,  \qquad
\delta g_{\mu\nu} = 2 \epsilon (x) g_{\mu\nu},
\label{Comp-Weyl}  
\end{eqnarray}
and a local translation for the Nakanishi-Lautrup field $B$:
\begin{eqnarray}
\delta B = - \epsilon (x).
\label{B-local}  
\end{eqnarray}
Note that the translation for the dilaton in (\ref{Dilaton-Var}) corresponds to the Weyl transformation
for the scalar field in (\ref{Comp-Weyl}).  Under the infinitesimal transformations (\ref{OSp-rot}),
(\ref{Comp-Weyl}) and (\ref{B-local}), the quantum Lagrangian is transformed as
\begin{eqnarray}
\delta {\cal L}_q &=& - \frac{1}{2} \sqrt{-g} g^{\mu\nu} \phi^2 \partial_\mu \varepsilon_{NM} 
X^M \overset{\leftrightarrow}{\partial}_\nu X^N - \frac{1}{2} \partial_\mu ( \sqrt{-g} g^{\mu\nu} \phi^2 
\partial_\nu \epsilon)
\nonumber\\
&=& - \frac{1}{2} \partial_\mu \varepsilon_{NM} {\cal M}^{\mu MN}
- \frac{1}{2} \partial_\mu ( \sqrt{-g} g^{\mu\nu} \phi^2 \partial_\nu \epsilon).
\label{OSp-current}  
\end{eqnarray}
Hence, the quantum Lagrangian is invariant under the $OSp$ rotation (\ref{OSp-rot}) up to a
surface term when an additional Weyl transfomation (\ref{Comp-Weyl}) and a local translation (\ref{B-local}) 
are achieved.

At this stage, we critically make use of our result obtained thus far:
There is no Noether current associated with the Weyl symmetry, which makes it possible to
neglect the last term on the right-hand side of Eq. (\ref{OSp-current}) in evaluating the current
for the $OSp$ rotation (\ref{OSp-rot}). Consequently, the current for the $OSp$ rotation (\ref{OSp-rot})
is given by ${\cal M}^{\mu MN}$. In a perfectly similar way, we can derive the conserved current 
${\cal P}^{\mu M}$ for the infinitesimal translation in (\ref{Cons-currents}) \cite{Oda-Q}:
\begin{eqnarray}
\delta X^M = \varepsilon^M.
\label{transl}  
\end{eqnarray}
In this way, by using the phenomenon of the vanishing Weyl Noether currents, we have succeeded 
in obtaining the currents of the choral symmetry, which provide us with the basic elements of 
all the global symmetries such as the BRST symmetry and the conformal symmetry
in the Weyl invariant quantum gravity \cite{Oda-Q}.

\section{Conclusion}

In this article, we have not only proved that the Noether current associated with the Weyl symmetry 
(or a local scale symmetry) is identically vanishing by means of the second Noether theorem, but also clarified 
its reason that the transformation law of the Weyl transformation does not contain the derivative term
of a transformation parameter.

Moreover, we have applied this result to a quantum theory of a general Weyl invariant gravity
and derived the Noether currents associated with choral symmetry, which is a global $IOSp(10|10)$ supersymmetry.
In this derivation, the absence of the Weyl Noether current plays an important role. It is of interest that
the existence of the Noether currents usually provides us with useful information on the structure of symmetries
in a theory while in case of the Weyl symmetry the absence of the Noether current gives us useful knowledge
on the other symmetries. This situation reminds us of a precept of Buddhism, ``Every form in reality is empty,
and emptiness is the true form.''

As future's works, we wish to complete a quantization of a Weyl invariant gravity within the BRST formalism
\cite{Oda-Q, Oda-D} and investigate the problem of confinement of massive ghost \`a la BRST, for which the choral symmetry,
which is constructed in this article owing to the absence of the Weyl current, would become an essential ingredient. 
We also wish to construct a Weyl invariant regularization scheme by introducing an additional scalar field 
which plays a role of the renormalization mass scale $\mu$, by which the Weyl symmetry would not be violated 
by anomalies but be spontaneously broken.

\section*{Acknowledgment}

This work is supported in part by the JSPS Kakenhi Grant No. 21K03539.


\end{document}